\documentclass[prd,aps,superscriptaddress,nofootinbib,showpacs,12pt]{revtex4-1}
\DeclareRobustCommand{\baselinestretch{1.33}}

\usepackage{amsmath,amssymb,amsbsy}
\usepackage{enumerate}

\def\bk{{\boldsymbol k}}
\def\bp{{\boldsymbol p}}
\def\br{{\boldsymbol r}}
\def\bx{{\boldsymbol x}}

\def\ri{{\rm i}}

\def\beq{\begin{equation}}
\def\eeq{\end{equation}}
\def\ber{\begin{eqnarray}}
\def\eer{\end{eqnarray}}

\def\eff{{\rm eff}}

\begin{document}

\title{Generalizing the Cosmic Energy Equation}

\author{Yuri Shtanov}
\affiliation{Bogolyubov Institute for Theoretical Physics, Kiev 03680, Ukraine}
\affiliation{Department of Physics, Taras Shevchenko Kiev National University, Kiev,
Ukraine}

\author{Varun Sahni}
\affiliation{Inter-University Centre for Astronomy and Astrophysics, Post Bag 4,
Ganeshkhind, Pune 411~007, India}

\begin{abstract}
We generalize the cosmic energy equation to the case when massive particles interact via
a modified gravitational potential of the form $\phi (a, r )$, which is allowed to
explicitly depend on the cosmological time through the expansion factor $a(t)$. Using the
nonrelativistic approximation for particle dynamics, we derive the equation for the
cosmological expansion which has the form of the Friedmann equation with a renormalized
gravitational constant. The generalized Layzer--Irvine cosmic energy equation and the
associated cosmic virial theorem are applied to some recently proposed modifications of
the Newtonian gravitational interaction between dark-matter particles. We also draw
attention to the possibility that the cosmic energy equation may be used to probe the
expansion history of the universe thereby throwing light on the nature of dark matter and
dark energy.
\end{abstract}


\maketitle

\section{Introduction}

Almost half a century ago, Layzer \& Irvine \cite{LI}, and independently Zeldovich \&
Dmitriev \cite{zel}, demonstrated that in an expanding universe, the {\em peculiar\/}
kinetic ($K$) and potential ($U$) energies of a large system of pressureless particles
interacting via the Newtonian potential $\phi \equiv - G/r$ satisfy the {\em cosmic
energy equation}
\beq \label{eq:energy}
{\dot E} = - (2K + U) H \, , \qquad E = K + U \, ,
\eeq
where (see also \cite{peebles1,peebles2})
\beq \label{eq:k}
K = \frac12 \sum_i m_i v_i^2 \, ,
\eeq
and
\beq \label{eq:u}
 U = - \frac{G}{2} \int \frac{ \left[ \rho (\br_1) -
\varrho \right] \left[ \rho (\br_2) - \varrho \right] }{| \br_1 - \br_2 |}\, d^3 \br_1
d^3 \br_2
\eeq
are, respectively, the peculiar kinetic and potential energies of a system of particles
$m_i$ having coordinates $\br_i$ and peculiar velocities $v_i$.  Here, $\rho (\br)$ is
the mass density of these particles, $\varrho$  is its cosmological background average,
and $H = \dot a / a$ is the Hubble parameter describing the cosmic expansion.  For the
discrete distribution $\rho (\br) = \sum_i m_i \delta (\br - \br_i)$, the integral in
(\ref{eq:u}) should avoid the configuration $\br_1 = \br_2$.

One can introduce the correlation function $\xi (r)$:
\beq \label{eq:corr}
\left\langle \left[\rho (\br_0 + \br) - \varrho \right] \left[\rho (\br_0) - \varrho
\right] \right\rangle = \varrho^2 \xi (r)
\eeq
with the obvious property
\beq \label{eq:property}
\int \xi (r)\, d^3 \br = 4 \pi \int \xi (r)\, r^2 dr = 0 \, .
\eeq
Then Eq.~(\ref{eq:energy}) can be averaged to yield the corresponding equation per unit
mass
\beq \label{eq:unitm}
\langle \dot {\cal E}  \rangle = - \left[ 2 \langle {\cal K} \rangle + \langle {\cal U}
\rangle \right] H
\eeq
with ${\cal E} = E/M$, ${\cal U} = U/M$, and
\beq \label{eq:uaver}
\langle {\cal U} \rangle  
= - 2 \pi G \varrho \int \xi (r) r dr \, .
\eeq
Here, $M = \varrho V$ is the total mass of the system in a large volume $V$.

Two important limiting cases of Eq.~(\ref{eq:energy}) need mention
\cite{peebles1,peebles2}:
\begin{enumerate}[(i)]

\item For noninteracting particles, we have $U = 0$, and the relation ${\dot K} = - 2
    H K$ corresponds to the kinematic decay of the peculiar velocities with time $v_i
    \propto 1/a(t)$.

\item Once the clustering has entered a stationary regime, the condition ${\dot E} =
    0$ results in the virial relation
\beq \label{eq:virial1}
2  K + U = 0 \, .
\eeq

\end{enumerate}

A breakdown of (\ref{eq:virial1}) for galaxies belonging to the Coma cluster led Zwicky
\cite{zwicky} to suggest that a large amount of dark matter might dominate the dynamics
of Coma. While dark matter appears to be ubiquitous, its nature remains elusive. Indeed
it is now believed that, to properly account for the energy budget of the universe, dark
matter (DM) must be supplemented by an even more enigmatic `substance' called dark energy
(DE), which, on account of its large negative pressure, causes the universe to accelerate
instead of decelerating.

Our current theoretical understanding of DM and DE can be broadly divided into ideas that
are mainstream and those that are radical. Mainstream notions suggest that DM is formed
of nonbaryonic particles of relic origin while DE consists of the cosmological constant
or a relic scalar field such as quintessence. Radical notions suggest that the purported
existence of DM/DE may be pointing to a breakdown of Newtonian/Einsteinian gravity on
large scales. Theoretical models which incorporate this latter set of ideas include
Braneworld and $f(R)$ gravity theories in the case of DE, and Modified Newtonian Dynamics
(MOND) and the screened gravitational interaction model in the case of DM (see also
\cite{amendola,kamion}).

Ever since its discovery, the cosmic energy equation (\ref{eq:energy}) has been one of
the bulwarks of modern cosmology, and its many applications include estimates of the
matter density and its gravitational binding energy \cite{peebles1,peebles2,fukupeeb}. In
this paper, we show how the cosmic energy equation can be generalized to incorporate more
flexible forms of the gravitational interaction of dark matter, some of which have been
suggested in the literature.

Our assumption is that nonrelativistic dark-matter particles interact with each other via
the two-particle potential $\phi (a, r)$ so that the potential energy between two
particles is
\beq \label{eq:pot}
m_1 m_2 \phi \left(a, | \br_1 - \br_2 | \right) \, .
\eeq
As indicated in (\ref{eq:pot}), we allow the two-particle potential to depend explicitly
on the cosmological time through the scale factor $a(t)$, which is characteristic of some
class of the models to be considered below. We consider the universe dominated by these
dark-matter particles and by the uniform dark energy.

In Sec.~\ref{sec:cosmology}, using the nonrelativistic theory, we obtain the law of
cosmological expansion for our universe, which turns out to have the Friedmannian form
with a renormalized gravitational constant.  In Sec.~\ref{sec:LI}, we then derive the
generalized cosmic energy equation and the corresponding virial relations, which are
analogs of the corresponding Layzer--Irvine equation (\ref{eq:energy}) and virial
relation (\ref{eq:virial1}).  In Sec.~\ref{sec:discuss}, we comment on the results
obtained.

\section{Cosmological expansion}
\label{sec:cosmology}

In this section, we will derive the law of cosmological expansion given the interaction
potential (\ref{eq:pot}).  The universe model can be taken to be spatially flat since we
are dealing with spatial scales much smaller than the Hubble length.  We describe
particle dynamics within a large volume using the nonrelativistic approximation with
respect to comoving particle velocities.

\subsection{Expansion law in the $\Lambda$CDM model}

The Lagrangian for a nonrelativistic particle is given by the expression \cite{peebles1}
\beq \label{eq:one}
{\cal L} = \frac12 m \dot \br^2 - m \Phi (\br, t) \, ,
\eeq
where $\Phi (\br, t)$ is the full potential acting on the particle.  In the $\Lambda$CDM
(cosmological constant $\Lambda$ + cold dark matter) model, in which dark-matter
particles interact via Newtonian potential, for a pointlike matter distribution $\rho
(\br, t) = \sum_i m_i \delta (\br - \br_i)$, the potential $\Phi (\br , t)$ 
is given by
\beq \label{eq:phi1}
\Phi (\br , t) = - G \sum_j \frac{m_j}{|\br - \br_j|} - \frac{\Lambda}{6} r^2 \, .
\eeq

Proceeding to the comoving coordinates $\bx = \br / a$ and making a canonical
transformation
\beq
{\cal L} \to {\cal L} - \frac{d}{dt} \left( \frac12 m a \dot a x^2 \right) \, ,
\eeq
one transforms the one-particle Lagrangian (\ref{eq:one}) to
\beq \label{eq:one1}
{\cal L} = \frac12 m a^2 \dot \bx^2 - m \varphi \, ,
\eeq
where the new potential is given by
\beq
\varphi = \Phi + \frac12 a \ddot a x^2 \, .
\eeq
In the Newtonian case, with account taken of (\ref{eq:phi1}), the new potential is
\beq \label{eq:phi2}
\varphi = - \frac{G}{a} \sum_j \frac{m_j}{|\bx - \bx_j|} + \frac12 \left(a \ddot a -
\frac{\Lambda a^2}{3} \right) x^2 \, .
\eeq

The law of cosmological expansion can be derived directly from (\ref{eq:phi2}) under the
requirement that the potential $\varphi $ should not exert force on a test particle in
the limit of a homogeneous distribution of matter $\rho (\br) \equiv \varrho$. To see
this, we note that, in particular, the Laplacian of the potential should vanish in this
limit. Now, we have
\ber
\nabla^2 \varphi &=& 4 \pi G a^2 \sum_j m_j\, \delta (\br -
\br_j ) + 3 a \ddot a - \Lambda a^2 \nonumber \\
&\to& 4 \pi G a^2 \varrho + 3 a \ddot a - \Lambda a^2 \, .
\eer
Whence, we obtain one of the Friedmann laws of expansion
\beq \label{eq:fr1}
\frac{\ddot a}{a} =  - \frac{4 \pi G}{3} \varrho + \frac{\Lambda}{3} \, .
\eeq

\subsection{Expansion law in the case of general potential}
\label{sec:Expansion}

In the case of a general two-particle potential (\ref{eq:pot}), the derivation that led
to the cosmological potential (\ref{eq:phi2}) for a particle can be repeated and
generalized to
\beq \label{eq:phi3}
\varphi  =  \sum_j m_j \phi (a, a |\bx - \bx_j| ) + \frac12 \left(a \ddot a -
\frac{\Lambda_\eff a^2}{3} \right) x^2 \, ,
\eeq
where $\Lambda_\eff$ is the time-dependent ingredient in the universe which describes
dark energy and replaces the cosmological constant and, in particular, whose interaction
with dark matter makes the potential (\ref{eq:pot}) time-dependent. Our equations below
will be general and will not require the specific knowledge of this term; we will only
assume that it does not cluster, remaining homogeneous in space.  One such
field-theoretic example is considered in Sec.~\ref{sec:model} below (example
\ref{example}).

The class of theories under consideration in this paper could, in fact, be defined by the
property that the potential acting on a single dark-matter particle has the form
(\ref{eq:phi3}). This complicated potential already includes the response of the
universe, with all its field-theoretic ingredients, to the presence of dark-matter
particles at specified spatial positions.  In this subsection, we are going to derive the
modified expansion law by taking the Laplacian of (\ref{eq:phi3}) and demanding that it
vanish in the case of a spatially uniform distribution of matter.

Suppose that the gravitational two-particle potential has the form (we suppress the
dependence on $a$ as it is not relevant here)
\beq \label{eq:phif}
\phi (r) = - \frac{G}{r} f (r) \, ,
\eeq
with a sufficiently regular function $f(r)$ [in particular, $\displaystyle \lim_{r \to 0}
r f'(r) = 0$], as will be the case in all our concrete examples. Then, calculating its
Laplacian, we have
\ber
&&\!\!\!\!\! \nabla_\br^2 \phi (r) = - G f(r) \nabla_\br^2 \frac{1}{r} - 2 G  \nabla_\br
\frac{1}{r} \cdot \nabla_\br f(r) - \frac{G}{r} \nabla_\br^2 f(r) \nonumber \\
&& = 4 \pi G f(0)\, \delta (\br) + \frac{2 G}{r^2} f'(r) - \frac{G}{r} \left[ f'' (r) +
\frac{2}{r} f'(r) \right] \nonumber \\
&& = 4 \pi G f(0)\, \delta (\br) - \frac{G}{r} f'' (r) \, .
\eer
Applying this to (\ref{eq:phi3}) and requiring that $\nabla^2 \varphi \equiv 0$ for
uniformly distributed matter, we obtain the equation
\beq \label{eq:fr2}
\frac{\ddot a}{a} = - \frac{4 \pi G_\eff}{3} \varrho + \frac{\Lambda_\eff}{3} \, ,
\eeq
where the effective gravitational constant is given by
\beq \label{eq:geff}
G_\eff = G \lim_{r \to \infty} \left[ f(r) - r f'(r) \right] \, .
\eeq
The derivation works as long as the limit in (\ref{eq:geff}) exists and is finite, which
we assume to be the case.  For the Newtonian potential $\phi = - G/r$, we have $f \equiv
1$, and Eq.~(\ref{eq:geff}) returns the usual gravitational constant $G$. However, if the
potential decays faster than $1/r$ at large distances, Eq.~(\ref{eq:geff}) may imply
$G_\eff = 0$, so that the gravity of matter will not influence the rate of cosmic
expansion. Note that the effective gravitational constant will be time-dependent if the
function $f$ in  (\ref{eq:phif}) depends on time.

\subsection{Examples}
\label{sec:model}

To give examples to which our results can be applied, we consider some simple
modifications of the gravitational interaction between dark-matter particles:
\begin{enumerate}[(a)]

\item \label{example} The screened potential discussed in \cite{nusser} has the form
\beq \label{eq:screen}
\phi (a, r) = - \frac{G}{r} \left( 1 + \beta e^{- r / r_s } \right) \, ,
\eeq
where $\beta$ is a dimensionless constant of order unity, and the time-dependent
screening length $r_s (t)$ is a comoving constant, i.e., $r_s (t) \propto a (t) $, such
that $r_s (t_0) \simeq 1$~Mpc.  The potential arises in the field-theoretic model of
interaction of dark matter with dark energy via the scalar field \cite{Farrar:2003uw} in
the version with two dark-matter families.  A subdominant relativistic family is used to
stabilize the value of the scalar field; then the dominant nonrelativistic dark-matter
particles have constant mass and interact via gravity as well as via the scalar field so
that the two-particle interaction potential (\ref{eq:screen}) is generated.  A
dark-matter particle then moves in potential (\ref{eq:phi3}) with $\phi$ given by
(\ref{eq:screen}), so that our analysis is applicable to this situation.  One arrives at
the Friedmann equation (\ref{eq:fr2}) with constant dark-energy term $\Lambda_\eff$, just
as in the field-theoretic approach of \cite{nusser,Farrar:2003uw}. The potential
(\ref{eq:screen}) leads to faster evacuation of matter from voids and to an earlier epoch
of structure formation, which could be perceived as an advantage for this model over
$\Lambda$CDM \cite{nusser}.

Using the potential (\ref{eq:phi3}) and the equations of motion for a dark-matter
particle $\bx (t)$,
\beq \label{eq:mot}
\ddot \bx + 2 H \dot \bx = - \frac{1}{a^2} \nabla \varphi \, ,
\eeq
and proceeding along the same lines as in \cite{peebles1}, Sec.~27, one easily obtains
the exact nonlinear evolution equation for the Fourier components $\delta_\bk$ of the
density contrast $\rho (\bx)/ \varrho - 1$ in this theory:
\beq \label{eq:nonlin}
\ddot \delta_\bk + 2 H \dot \delta_\bk = 4 \pi G \varrho \left[ 1 + \frac{\beta}{1 +
\left(a/ k r_s \right)^{2}} \right] \delta_\bk + A_\bk - C_\bk \, ,
\eeq
where the nonlinear term $A_\bk$ and the velocity term $C_\bk$ are given by
\beq
A_\bk = 4 \pi G \varrho \sum_{\bk' \ne 0, \bk} \left[ \frac{\bk \bk'}{k'^2} + \frac{\beta
\bk \bk'}{k'^2 + \left(a/r_s \right)^2} \right] \delta_{\bk - \bk'} \delta_{\bk'}
\eeq
and
\beq
C_\bk = \sum_i \frac{ m_i}{M} \left( \bk \dot \bx_i \right)^2 e^{\ri \bk \bx_i} \, ,
\eeq
respectively.  The linear part of (\ref{eq:nonlin}) reproduces equation (6) of
\cite{nusser} and leads to a more rapid development of gravitational instability than in
$\Lambda$CDM.

\item A power-law correction to the Newtonian potential on large scales
\beq \label{eq:power}
\phi(r) = -\frac{G}{r} \left[ 1 + \left( \frac{r_0}{r + r_0} \right)^n \right]\, ,
\qquad n \ge 1 \, ,
\eeq
which we have regularized on small scales to avoid a singularity.  Potentials
 of this
form with $n = 2$ arise in the Randall--Sundrum model \cite{rs99} with a single large
extra dimension. (However, in this case, $r_0 \ll 1$~mm, making  the correction on
cosmological scales extremely small.)

\item A logarithmic correction to the Newtonian potential
\beq \label{eq:log}
\phi(r) = -\frac{G}{r} + \left (\frac{\alpha G}{r_0} \log \frac{r}{r_0} \right ) e^{-
r/r_c} \, .
\eeq
The influence of the `regularizing' exponent is confined to very large scales $r_c
\gg r_0$.

\end{enumerate}

For all these potentials the limit (\ref{eq:geff}) gives
\beq
G_\eff = G \, ,
\eeq
implying that the background cosmological evolution remains unmodified.

\section{Generalized Layzer--Irvine equation}
\label{sec:LI}

The Lagrangian of the many particle  system is obtained by the summation of
(\ref{eq:one1}) with account taken of (\ref{eq:phi3}), (\ref{eq:fr2}) and (\ref{eq:geff})
and is equal to $L = K - U$, where the peculiar kinetic energy is
\cite{peebles1,peebles2}
\beq \label{eq:K}
K = \frac{a^2}{2} \sum_i m_i \dot x_i^2 \, .
\eeq
With consideration of Eq.~(\ref{eq:fr2}), potential (\ref{eq:phi3}) can be written in the
form
\beq
\varphi = \int \left[ \rho(\br') - \varrho \right] \phi (a, |\br - \br'|)\, d^3 \br' \, ,
\eeq
using which, one obtains
\beq \label{eq:U1}
U = \frac12 \int \left[ \rho (\br_1) - \varrho \right] \left[ \rho (\br_2) - \varrho
\right] \phi (a, r_{12}) \, d^3 \br_1 d^3 \br_2 \, ,
\eeq
where, for the discrete distribution $\rho (\br) = \sum_i m_i \delta (\br - \br_i)$, the
integral should avoid the configuration $\br_1 = \br_2$.  Here and below, we use the
notation $r_{ij} = |\br_i - \br_j |$ and $x_{ij} = |\bx_i - \bx_j |$.

The Hamiltonian of our system as a function of the canonical variables $(\bx_i, \bp_i )$
is given by
\beq
{\cal H} = K + U = \frac{1}{2 a^2} \sum_i \frac{p_i^2}{m_i} + \frac12 \int \left[ \rho
(\bx_1) - \varrho \right] \left[ \rho (\bx_2) - \varrho \right] a^6 \phi (a, a x_{12}) \,
d^3 \bx_1 d^3 \bx_2 \, .
\eeq
Using the fact that the product $a^3 \left[ \rho (\br) - \varrho \right] = \sum_i m_i
\delta (\bx - \bx_i) - a^3 \varrho$ does not depend explicitly on time, we immediately
obtain the equation for the peculiar energy of the system:
\ber \label{li}
{\dot E} &\equiv& \frac{d}{dt} \left( K + U \right) = \frac{\partial {\cal H}}{\partial
t} \nonumber \\ &=& - 2 H K + \frac{H}{2} \int \left[ \rho (\br_1) - \varrho \right]
\left[ \rho (\br_2) - \varrho \right] \left[ \frac{\partial \phi \left(a, r_{12}
\right)}{\partial r} \, r_{12} + \frac{\partial \phi \left(a,  r_{12} \right)}{\partial
a}\, a \right] d^3 \br_1 d^3 \br_2  \, . \quad
\eer

Similarly to (\ref{eq:unitm}) and (\ref{eq:uaver}), we can write the averaged equations
for quantities per unit mass:
\beq \label{eq:li}
\langle \dot {\cal E} \rangle = - 2 H \langle {\cal K} \rangle + 2 \pi H \varrho \int \xi
(r) \left[ \frac{\partial \phi \left(a, r \right)}{\partial r} \, r + \frac{\partial \phi
\left(a,  r \right)}{\partial a}\, a \right] r^2 dr \, ,
\eeq
\beq \label{eq:uaver1}
\langle {\cal U} \rangle  = \frac{\varrho^2}{2 M} \int \xi (|\br_1 - \br_2 |) \phi (a, |
\br_1 - \br_2 |) \, d^3 \br_1 d^3 \br_2 = 2 \pi \varrho \int \xi (r) \phi (a, r)\, r^2 dr
\, .
\eeq
Equations (\ref{li})--(\ref{eq:uaver1}) form the main results of this letter.

In the Newtonian case, we have $\partial \phi / \partial r = - \phi / r$,  $\partial \phi
/ \partial a = 0$, and Eqs.~(\ref{li}) and (\ref{eq:li}) reduce to the corresponding
Layzer--Irvine Eqs.~(\ref{eq:energy}) and (\ref{eq:unitm}). In the general case,
Eqs.~(\ref{li}) and (\ref{eq:li}) contain an unknown function $\phi (a, r)$. If one has a
theory predicting the shape of $\phi$, one can, in principle, use (\ref{li}) and
(\ref{eq:li}) to test it.

Once the system has decoupled from the Hubble expansion, its peculiar energy evolves
mainly because of the time-dependence of the potential $\phi (a, r)$:
\beq \label{eq:envir}
\langle \dot {\cal E} \rangle \approx 2 \pi H \varrho \int \xi (r) \frac{\partial \phi
\left(a, r \right)}{\partial a}\, a r^2 dr \, .
\eeq
Taking into account (\ref{eq:li}), we obtain the generalized virial relation in the form
\beq \label{eq:genvir}
\langle {\cal K} \rangle = \pi \varrho \int \xi (r) \frac{\partial \phi \left(a, r
\right)}{\partial r} \, r^3 dr \, .
\eeq

Next, we apply the generalized cosmic energy equation to the modifications of the
gravitational interaction between dark-matter particles that we listed in
Sec.~\ref{sec:model}:
\begin{enumerate}[(a)]

\item Substituting the screened potential (\ref{eq:screen}) into (\ref{li}), one
    finds, quite remarkably, that the generalized cosmic energy equation reduces to
    its Layzer--Irvine form (\ref{eq:energy}) with the {\em modified\/} total
    potential $U$ determined by (\ref{eq:screen}) and (\ref{eq:U1}). For a system
    decoupled from the Hubble expansion, the energy evolution, according to
    (\ref{eq:envir}), is given by
\beq
\langle \dot {\cal E} \rangle = - \frac{2 \pi \beta G H \varrho}{r_s} \int \xi (r)
 e^{- r/r_s}  r^2 dr \, ,
\eeq
and the generalized virial relation (\ref{eq:genvir}) in this case reads
\beq \label{screenvir}
2 \langle {\cal K} \rangle + \langle {\cal U} \rangle = \frac{2 \pi \beta G
\varrho}{r_s} \int \xi (r) e^{- r/r_s}  r^2 dr \, ,
\eeq

On length scales $r \ll r_s (t) \leq 1$~Mpc, we recover the usual Newtonian
relations:
\beq \label{correction}
- \frac{\langle \dot {\cal E} \rangle}{H} = 2 \langle {\cal K} \rangle + \langle {\cal U}
\rangle_{\rm Newton}  = \frac{2 \pi \beta G \varrho}{r_s} \int \xi (r) r^2 dr = 0 \, ,
\eeq
where, in the last equality, we have used property (\ref{eq:property}), and
\beq \label{asymp}
\langle {\cal U} \rangle_{\rm Newton} = - 2 \pi (1 + \beta)  G \varrho \int \xi (r) r
dr
\eeq
is the Newtonian virial gravitational energy per unit mass with renormalized
gravitational coupling.

It should be noted that, since $r_c \simeq 1$~Mpc roughly corresponds to the Abell
radius associated with a cluster of galaxies, it is unlikely that (\ref{correction})
can be applied to galaxy clusters. On these scales, the full form of the virial
relation (\ref{screenvir}) should be used.

\item The power-law correction to the Newtonian potential (\ref{eq:power}) leads to
    the cosmic energy equation:
\beq
\langle \dot {\cal E} \rangle \approx - \left[ 2 \langle {\cal K} \rangle + \langle
{\cal U} \rangle_{\rm Newton} \right] H + 2 \pi (1 + n) G H \varrho \int \xi (r)
\left( \frac{r_0}{r + r_0} \right)^n r dr \, ,
\eeq
and to the virial relation
\beq
2 \langle {\cal K} \rangle + \langle {\cal U} \rangle_{\rm Newton} = 2 \pi (1 + n) G H
\varrho \int \xi (r) \left( \frac{r_0}{r + r_0} \right)^n r dr \, ,
\eeq
where $\langle {\cal U} \rangle_{\rm Newton}$ is the usual averaged Newtonian
peculiar potential energy per unit mass given by (\ref{eq:uaver}).

\item The logarithmic correction to the Newtonian potential (\ref{eq:log}) leads to
    the usual cosmic energy equation
\beq
\langle \dot {\cal E} \rangle \approx - \left[ 2 \langle {\cal K} \rangle + \langle
{\cal U} \rangle_{\rm Newton} \right] H + \frac{2 \pi \alpha G H \varrho}{r_0} \int
\xi (r) r^2 e^{-r/r_c} dr \approx - \left[ 2 \langle {\cal K} \rangle + \langle {\cal
U} \rangle_{\rm Newton} \right] H \, ,
\eeq
where the last equality is valid in view of (\ref{eq:property}) if the correlation
length of the system is much smaller than the cutoff scale $r_c$.  The virial
relation in this case is
\beq \label{virial:log}
2 \langle {\cal K} \rangle + \langle {\cal U} \rangle_{\rm Newton} = \frac{2 \pi \alpha G
\varrho}{r_0} \int \xi (r) r^2 e^{-r/r_c} dr \approx 0 \, .
\eeq

\end{enumerate}

\section{Discussion}
\label{sec:discuss}

We have investigated a cosmological theory which produces a modified and, perhaps,
time-dependent gravitational potential between matter particles in the form
(\ref{eq:pot}).  For such a theory, we have determined the background cosmological
equation, which turned out to have the Friedmann form possibly with a modified
gravitational constant.  We have also derived the cosmic energy equation generalizing the
Layzer--Irvine equation for the theory under investigation.

We also note that the cosmic energy equation (\ref{eq:unitm}) or its generalization
(\ref{eq:li}) can be used to determine the cosmic expansion history. For instance,
from (\ref{eq:energy}) we obtain
\beq
H(t) = - \frac{\langle \dot {\cal E} \rangle}{2 \langle {\cal K} \rangle + \langle {\cal
U} \rangle} \, .
\eeq

The Hubble parameter $H(z)$ determined in this manner could shed light on the nature of
dark energy either through the $Om$ diagnostic \cite{om} or by means of the effective
equation of state of dark energy \cite{ss06}
\beq\label{eq:state}
w_{\rm DE}(x) = \frac{(2 x /3)\, d \ln H / dx  - 1}{1 - (H_0/H)^2\, \Omega_{m}\, x^3} \,
, \qquad x = 1 + z \, .
\eeq

This determination of the cosmic expansion history $H(z)$, via the energy equation, is
complementary to usual methods which rely either on standard candles (supernovae of type
Ia) or rulers (baryon acoustic oscillations). This possibility will be examined in detail
elsewhere.

\section*{Acknowledgments}

The work of Y.~S.\@ was supported by the Cosmomicrophysics programme and by the Program
of Fundamental Research of the Physics and Astronomy Division of the National Academy of
Sciences of Ukraine. The authors acknowledge useful correspondence with Alexei
Starobinsky.


\begin{thebibliography}{99}

\bibitem{LI}
D.~Layzer,  ApJ {\bf 138}, 174L (1963); W.~M.~Irvine, PhD dissertation, Harvard
University (1961).

\bibitem{zel}
N.~A.~Dmitriev and Ya.~B.~Zeldovich, Zh. Eksp. Teor. Fiz. {\bf 45}, 1150 (1963) [Sov.
Phys. JETP {\bf 18}, 793 (1963)].

\bibitem{peebles1}
P.~J.~E.~Peebles, {\em The Large-Scale Structure of the Universe}, Princeton University
Press, Princeton (1980).

\bibitem{peebles2}
P.~J.~E.~Peebles, {\em Principles of Physical Cosmology}, Princeton University Press,
Princeton (1993).

\bibitem{zwicky}
F.~Zwicky, Helv. Phys. Acta {\bf 6}, 110 (1933).

\bibitem{amendola}
L.~Amendola and D.~Tocchini-Valentini,
  Phys.\ Rev.\  D {\bf 66}, 043528 (2002);
L.~Amendola and C.~Quercellini,
  Phys.\ Rev.\  D {\bf 68}, 023514 (2003).

\bibitem{kamion}
M.~Kesden and M.~Kamionkowski, \prd {\bf 74}, 083007 (2006).

\bibitem{fukupeeb}
M.~Fukugita and P.~J.~E.~Peebles, ApJ {\bf 616}, 643 (2004).

\bibitem{nusser}
A.~Nusser, S.~S.~Gubser and P.~J.~E.~Peebles,
Phys.\ Rev.\  D {\bf 71}, 083505 (2005).

\bibitem{Farrar:2003uw}
G.~R.~Farrar and P.~J.~E.~Peebles,
Astrophys.\ J.\  {\bf 604}, 1 (2004).

\bibitem{rs99}
L.~Randall and R.~Sundrum, Phys. Rev.  Lett. {\bf 83}, 3370 (1999); \ {\bf 83}, 4690
(1999); P.~Callin and F.~Ravndal, Phys. Rev. D {\bf 70}, 104009 (2004).

\bibitem{om}
V.~Sahni, A.~Shafieloo and A.~A.~Starobinsky, \prd {\bf 78}, 103502 (2008).

\bibitem{ss06}
V.~Sahni and A.~A.~Starobinsky, IJMP(D) {\bf 15}, 2105 (2006).
\end{thebibliography}
\end{document}